\documentclass[12pt]{iopart}

\usepackage{iopams}
\usepackage{graphicx}
\usepackage{amssymb}
\usepackage{color}

\usepackage[colorlinks, linkcolor={blue},
    citecolor={blue}, urlcolor={blue}]{hyperref}

\newcommand{\kp}{k_{\parallel}}
\renewcommand{\k}{\boldsymbol{k}}
\renewcommand{\a}{\boldsymbol{a}}
\renewcommand{\r}{\boldsymbol{r}}

\begin{document}

\title[Manipulation of edge states in microwave artificial graphene]{Manipulation of edge states in microwave artificial graphene}

\author{Matthieu Bellec,$^1$ Ulrich Kuhl,$^1$ Gilles Montambaux,$^2$ and Fabrice Mortessagne$^1$}
\address{$^1$ Laboratoire Physique de la Mati\`{e}re Condens\'{e}e, CNRS UMR 7336, Universit\'{e} Nice-Sophia Antipolis, 06100 Nice, France}
\address{$^2$ Laboratoire de Physique des Solides, CNRS UMR 8502, Universit\'{e} Paris-Sud, 91405 Orsay Cedex, France}

\ead{fabrice.mortessagne@unice.fr}

\begin{abstract} 
Edge states are one important ingredient to understand transport properties of graphene nanoribbons. We study experimentally the existence and the internal structure of edge states under uniaxial strain of the three main edges: zigzag, bearded, and armchair. The experiments are performed on artificial microwave graphene flakes, where the wavefunctions are obtained by direct imaging. We show that uniaxial strain can be used to manipulate the edge states: a single parameter controls their existence and their spatial extension into the ribbon. By combining tight-binding approach and topological arguments, we provide an accurate description of our experimental findings. A new type of zero-energy state appearing at the intersection of two edges, namely the corner state, is also observed and discussed.
\end{abstract}

\hspace*{1.5cm}
\parbox{13cm}{{\it Keywords}: artificial graphene, edge states, photonics crystal, microwave experiments\\[2ex]
Published 11 November 2014 in New Journal of Physics 16 (2014) 113023}
\maketitle

\section{Introduction}

Edge states play an essential role in condensed matter physics for both fundamental aspects and electronic transport applications. Recently, due to their immunity against disorder or impurities (absence of backscattering), topologically protected edge states have raised interest in quantum spin Hall systems~\cite{Kane05, Bernevig06a, Bernevig06b} as well as in topological insulators~\cite{Haldane88, Hasan10}. Edge states were first predicted in graphene ribbons~\cite{Nakada96, Brey06, Kohmoto07} and later observed along ``zigzag'' edges~\cite{Kobayashi05, Niimi06}. Although they are not strictly speaking topologically protected, edge states in graphene possess a topological origin coined by the Zak phase and remain robust against weak chiral symmetry perturbations~\cite{Ryu02, Delplace11}.

These peculiar features are due to the multicomponent (spinorial) structure of the wavefunction. While for scalar wavefunctions (like for free particles in a box), hard wall boundary conditions impose the vanishing of the wavefunction at the edges, for a two-component wavefunction only one component has to vanish, leaving the possibility of a finite amplitude for the other component.
In graphene, the two components are simply the amplitudes of the wavefunction on each of the two atoms of the bipartite honeycomb lattice.

The richness of edge-state physics is not limited to condensed matter. Any finite system characterized by multi-component wavefunctions can constitute a good candidate. Over the past few years, pertinent realizations of artificial graphene have emerged in various contexts such as 2D electron gases in molecular assemblies~\cite{Gomes12} or in nanopatterned semiconductors~\cite{Singha11}, ultracold atoms in optical lattices~\cite{Tarruell12}, polaritons in semiconductor microcavities~\cite{Jacqmin14} as well as light and microwaves in photonic crystals~\cite{Rechtsman13a, Bittner10, Kuhl10, Bellec13a} (see~\cite{Polini13} for a recent review). The main advantages of these analogue systems lie in their high tunability and the control of their lattice properties. The synthetic honeycomb lattices offer the possibility to investigate phenomena that are hardly reachable in genuine graphene, particularly those appearing at the edges. For instance, edge states at bearded terminations (not stable in real graphene) have been first observed in photonic lattices~\cite{Plotnik14}. As pointed out in~\cite{Hafezi11, Fang12}, the manipulation and control of edge states may lead to promising photonic applications.

Moreover, graphene band structure, and consequently transport properties, can be engineered via lattice strain~\cite{Guinea10, Guinea12, Wunsch08, Montambaux09b, Pereira09}. While not observable in the realm of genuine graphene, topological phase transition with bandgap opening, as well as creation of a pseudo-magnetic field, have been observed in several artificial graphene realizations by implementing various strains of the honeycomb structure~\cite{Gomes12, Tarruell12, Rechtsman13a, Bellec13a, Rechtsman13b}.

\begin{figure}[t]
	\centering
	\includegraphics[width=16cm]{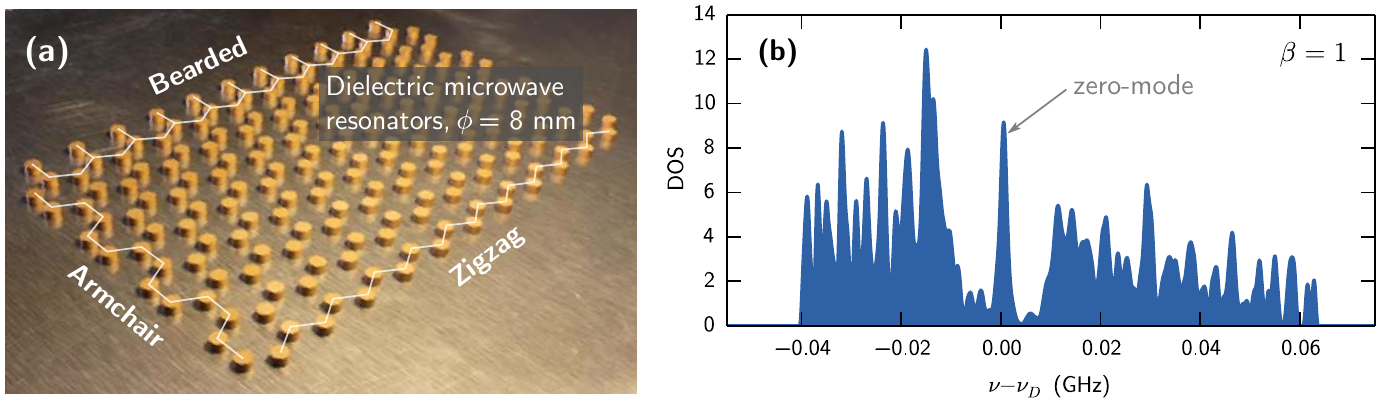}
	\caption{\label{Fig_Exp}(a) Picture of an unstrained artificial graphene ribbon with zigzag, bearded and armchair edges. The lattice constant is 15\,mm. (b) Corresponding experimental density of states (DOS). The arrow indicates the zero-modes appearing at the Dirac frequency $\nu_D$.}
\end{figure}

In this paper, we propose an experimental manipulation of edge state properties by controlling a uniaxial strain. We use a photonic implementation of honeycomb lattice in the microwave regime~\cite{Bellec13a, Bellec13b}. Figure~\ref{Fig_Exp}(a) shows a realization of an artificial graphene ribbon exhibiting the three usual types of edges, namely zigzag, bearded and armchair. The sites of the lattice are occupied by dielectric microwave resonators with a cylindrical shape (diameter of 8\,mm, height of 5\,mm). The resonance frequency of an isolated resonator $\nu_0$ is around $6.65$\,GHz and corresponds to the on-site energy of atoms in a tight-binding (TB) model. The dielectric cylinders are coupled by an evanescent magnetic field, so that the wave propagation between the resonators is well described by a tight-binding-like hopping term. Each resonator is mainly coupled to its three nearest neighbors. The coupling strength $t$ between two resonators depends on their separation $d$ and varies from $t=0.015$\,GHz to $t=0.053$\,GHz when $d$ varies from 15 to 11\,mm. Via a reflection measurement, we have access, at each site, to the local density of states and to the wavefunction intensity associated to each eigenfrequency. The density of states (DOS) is obtained by averaging the local density of states over all resonator positions. The experimental setup and the tight-binding description of the microwave artificial graphene are detailed in~\cite{Bellec13b}.

The paper is organized as follows. In section~\ref{Sec_BZstudy}, we first focus on zigzag and bearded boundary geometries. We show experimentally how uniaxial strain acts as a switch between zigzag and bearded edge states. Based on a tight-binding analysis, a diagram of the existence of edge states is theoretically proposed. We recall in section~\ref{Sec_Zak} the topological origin of the three types of considered edge states, namely zigzag, bearded and armchair. A geometrical analysis in the $k$-space allows to predict the presence of edge states and their evolution under strain. Section~\ref{Sec_ACstudy} is dedicated to armchair geometries. A quantitative experimental and theoretical analysis is done. The existence of a new type of state, appearing at the intersection of two type of edges, namely corner states, is eventually discussed.

\section{Zigzag and bearded edges in honeycomb lattice under uniaxial strain}
\label{Sec_BZstudy}

\begin{figure}[t]
	\centering
	\includegraphics[width=16cm]{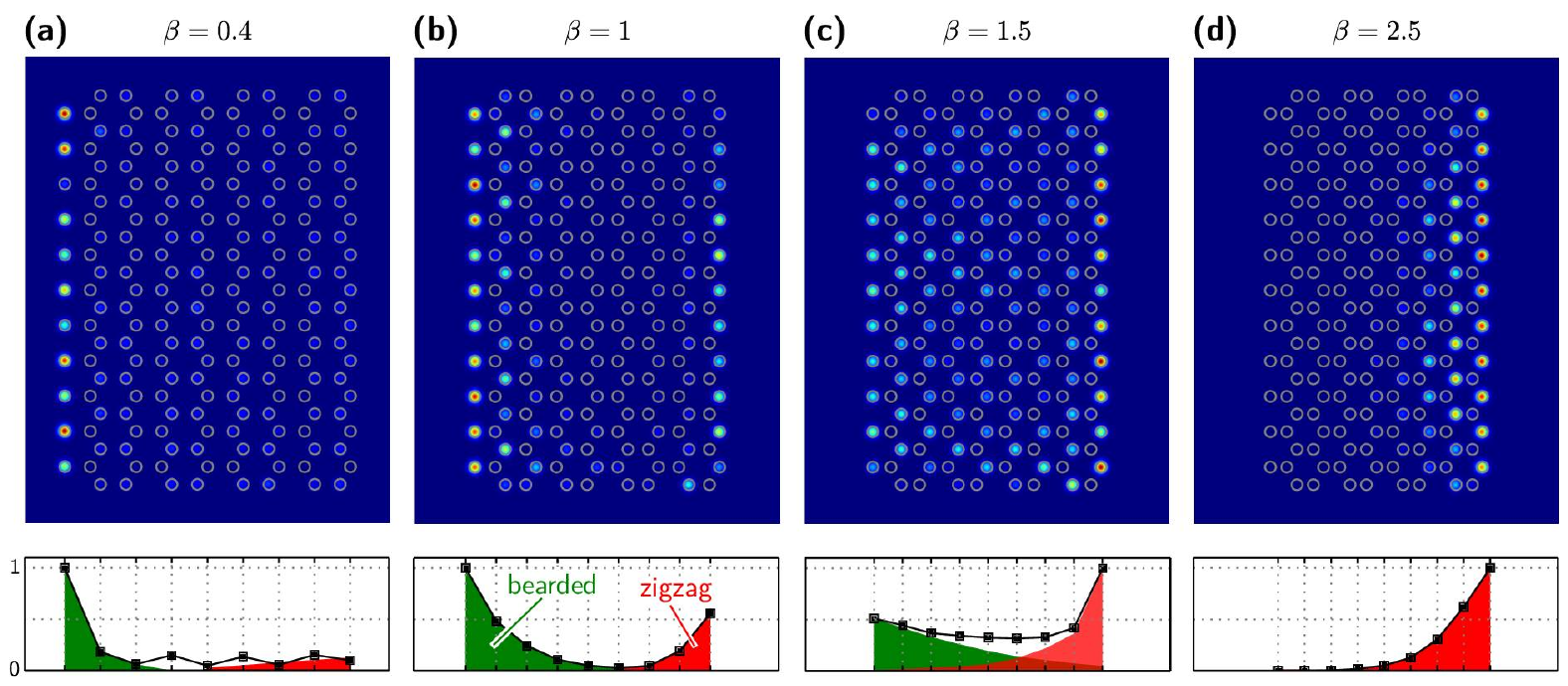}
	\caption{\label{Fig_WavefctBZ} (top row). Experimental zero-mode intensities for $\beta$ ranging from 0.4 to 2.5. The anisotropy axis is horizontal. Bearded and zigzag edges are displayed on the same lattice (respectively at the left and the right). (bottom row). Normalized intensities integrated over vertical lines. Green (resp. red) areas indicate qualitatively the bearded (resp. zigzag) zero-mode extension.}
\end{figure}

The lattice presented in figure~\ref{Fig_Exp}(a) exhibits three different edges: armchair, zigzag and bearded. We will consider ribbons uniaxially strained along one lattice axis (horizontal direction in figures~\ref{Fig_WavefctBZ} and~\ref{Fig_WavefctAC}) where the strain changes one of the three nearest-neighbor couplings only. The modified coupling is denoted by $t'$ and the anisotropy parameter by $\beta = t'/t$. Armchair edges are along the strain axis and consequently will not support any edge state whatever the anisotropy as will be discussed in section~\ref{Sec_Zak}.
Figure~\ref{Fig_Exp}(b) shows a typical DOS measured in an unstrained ribbon, i.e.\ $\beta = 1$. The Dirac frequency $\nu_D$ is obtained by following the procedure described in~\cite{Bellec13b} and defines the frequency origin. The peak observed at the origin corresponds to `zero-energy' modes in the condensed-matter context; we will call them `zero-modes' in the following. Experimentally, we can extract the intensity distributions by means of reflection measurements (see~\cite{Bellec13b} for details). Figure~\ref{Fig_WavefctBZ} shows the intensities of the wavefunctions associated to zero-modes for different values of $\beta$: the zero-modes are all located along edges. In the case of the unstrained lattice $\beta=1$, figure~\ref{Fig_WavefctBZ}(b), the intensity is clearly distributed along both zigzag and bearded boundaries. Then, the anisotropy parameter $\beta$ controls the relative weight between the two types of zero-modes. For $\beta = 0.4$ (figure~\ref{Fig_WavefctBZ}(a)), bearded edges are dominant whereas they are totally absent for $\beta = 2.5$ where only the zigzag edge is illuminated (figure~\ref{Fig_WavefctBZ}(d)). For intermediate anisotropy ($\beta = 1.5$, figure~\ref{Fig_WavefctBZ}(c)), both edge types are excited, with an opposite relative strength compared to the $\beta=1$ case.

\begin{figure}[t]
	\centering
	\includegraphics[width=16cm]{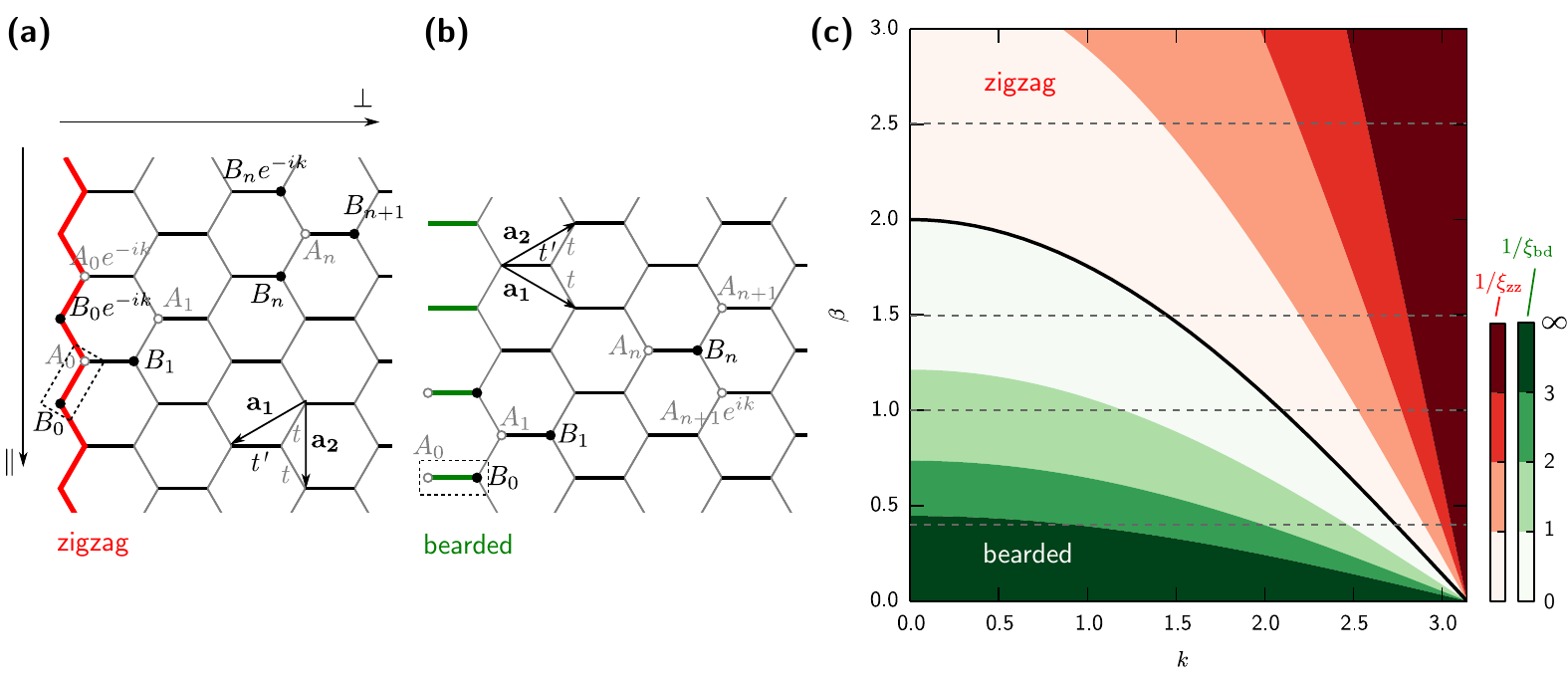}
	\caption{\label{Fig_DiagBZ} Schematic representation of a semi-infinite lattices with zigzag ((a), red) and bearded ((b), green) edge along the $\parallel$ axis. The anisotropy axis is horizontal, here along the $\perp$ axis. The corresponding coupling strength is denoted $t'$. $A_n$ and $B_n$ are the wavefunction amplitudes in the two sublattices. $k$ is the 1D Bloch wavevector (see text for details). (c) Zigzag -- bearded edge states existence diagram for various $k$ and $\beta$. The colormap corresponds to the inverse localization length in the zigzag (red, $1/\xi_{zz}$) and bearded (green, $1/\xi_{bd}$) cases; the darker the color, the stronger the states are localized. The dashed lines correspond to the values of $\beta$ used in the experiments.}
\end{figure}

To go beyond this qualitative discussion, we propose in the following part a theoretical description by means of a tight-binding model. Although both zigzag and bearded edges are built experimentally on the same ribbon sample, we will consider theoretically two independent semi-infinite lattices. We demonstrated in a previous work the validity of the TB model to describe our artificial graphene realization~\cite{Bellec13b}. Here, to address the issue of edge states, we will restrict the TB model to first nearest neighbor couplings.
Let us first consider a zigzag edge (figure~\ref{Fig_DiagBZ}(a)). The $A\!-\!B$ unit cell, appropriated to describe such an edge, is represented by the dashed box. The edge is built by translating the dimer with vectors $m\,\a_2$ ($m$ integer). According to the Bloch theorem, an additional phase $e^{i m k}$ is acquired after $m$ translations. Here, $k \equiv \sqrt{3} \kp a$, where $a$ is the lattice spacing and $\kp$ is the 1D wave vector pertaining to the edge. The 1D Brillouin zone (BZ) is defined by $\kp \in \left[ -\pi / \sqrt{3}a, \pi / \sqrt{3}a \right]$ corresponding to $k \in \left[ -\pi, \pi \right]$. For a given $m$, the bulk sites correspond to translations of the dimer with vectors $-n\,\a_1$ ($n$ positive integer). The presence of the edge breaks the translation symmetry and the Bloch theorem does not apply anymore. We label the site positions with the index $n$ from $n = 0$ (edge) to $n = N \rightarrow \infty$ (bulk) (see figure~\ref{Fig_DiagBZ}(a)).

Taking into account only nearest-neighbor couplings, each sublattice gives a tight-binding recurrence equation which defines the zero-modes~\cite{Kohmoto07}: $\sum_i t_i A_i = 0$ and $\sum_i t_i B_i = 0$, where $i$ counts the three nearest-neighbors of a given $B$-site and $A$-site, respectively, $t_i$ the corresponding coupling strength, and $A_i$ and $B_i$ denoting the amplitudes at the corresponding sites. In figure~\ref{Fig_DiagBZ}(a), the missing column of sites nearest to the edge belongs to the $A$ sublattice: $A_{-1}\equiv 0$. Thus, for zigzag edge, the tight-binding recurrence equation
\begin{equation}
\beta A_{n} + (1 + e^{i k}) A_{n+1} = 0 \label{Eq_A_BD}
\end{equation}
implies that all $A$-sites are identically null. For the $B$-sites, amplitudes of zero-modes fulfill the following condition:
\begin{equation}
\beta B_{n+1} + (1 + e^{-i k}) B_{n} = 0 \ . \label{Eq_B_ZZ}
\end{equation}
The amplitudes $B_n$ decay as
\begin{equation}
|B_n|^2 = \left( \frac{2}{\beta} \cos \frac{k}{2} \right)^{2n} \equiv e^{-n/\xi_{\mathrm{zz}}} \ , \label{Eq_Bn_ZZ}
\end{equation}
which defines a localization length $\xi_{\mathrm{zz}}(k, \beta)$ plotted with a red colorscale in figure~\ref{Fig_DiagBZ}(c). The zigzag edge states exist when
\begin{equation}
|k| > 2 \arccos \frac{\beta}{2} \ .
\label{Eq_existZZ}
\end{equation}
Conversely, for the bearded edges, sketched in figure~\ref{Fig_DiagBZ}(b), all $B$-sites are identically null.
For the $A$ sites, the recurrence relation is the same as above equation~(\ref{Eq_A_BD}) and the non-vanishing amplitudes $A_n$ decay as
\begin{equation}
|A_n|^2 = \left( \frac{2}{\beta} \cos \frac{k}{2} \right)^{-2n} \equiv e^{-n/\xi_{\mathrm{bd}}} \label{Eq_An_BD}
\end{equation}
and
\begin{equation}
|k| < 2 \arccos \frac{\beta}{2} \ .
\label{Eq_existBD}
\end{equation}

From equations~(\ref{Eq_existZZ}) and~(\ref{Eq_existBD}), we obtain the existence diagram of edge states depicted in figure~\ref{Fig_DiagBZ}(c). The red and green color scales give respectively the spatial extension of zigzag and bearded edge states extracted from equations~(\ref{Eq_Bn_ZZ}) and~(\ref{Eq_An_BD}).
For $\beta = 1$, the diagram shows that zigzag zero-modes occupy 1/3 of the 1D Brillouin zone whereas the proportion is 2/3 for the bearded states~\cite{Nakada96, Brey06, Kohmoto07,Delplace11}. Thus, the TB prediction for the ratio between number of bearded and zigzag states is 2. The diagram also shows that the zero-modes are more localized along zigzag edges than along bearded edges.

The square dots in figure~\ref{Fig_WavefctBZ} (bottom row) represent normalized measured intensities integrated over vertical lines of resonators. Green and red zones are used as a guide for the eyes and indicate the bulk extension of the bearded and zigzag zero-modes, respectively. The total intensity in each zone is proportional to the number of corresponding zero-modes \cite{Bellec13b}. For $\beta=1$, the measured ratio is 2.3, in close agreement with the expected value and the bearded edge states do indeed have a larger extension. For $\beta=0.4$, we observe a majority of strongly localized bearded edge states and only a few extended zigzag zero-modes. For $\beta=1.5$, both edge states are equally present, the zigzag ones being slightly more localized. Finally, for $\beta=2.5$, only zigzag zero-modes exist (the intensity along the bearded edge is strictly zero) with a larger extension compared to the previous case.
These observations are in good accordance with the features depicted in the diagram of figure~\ref{Fig_DiagBZ}(c) and demonstrate that the anisotropy parameter allows for an accurate manipulation of edges states.

In the next section, we propose a topological argument, first presented in~\cite{Ryu02, Delplace11}, to address the existence of zero-modes, not only for bearded and zigzag but also for armchair edges.

\section{Zak phase}
\label{Sec_Zak}

A simple geometrical way to describe the existence of edge states is to relate their existence to a topological quantity which is a 1D winding number called the Zak phase~\cite{Zak89}. The spectrum of an infinite ribbon of finite width containing $M$ dimers $A\!-\!B$ consists in $2M$ one-dimensional bands $\epsilon_j(k_\parallel)$. Therefore for each wave vector $k_\parallel$, there are $2M$ states. Among these $2M$ states, two zero-modes may be localized along the edges. We recall below the relation between their existence and the Zak phase~\cite{Ryu02, Delplace11}.

\begin{figure}[t]
	\centering
	\includegraphics[width=16cm]{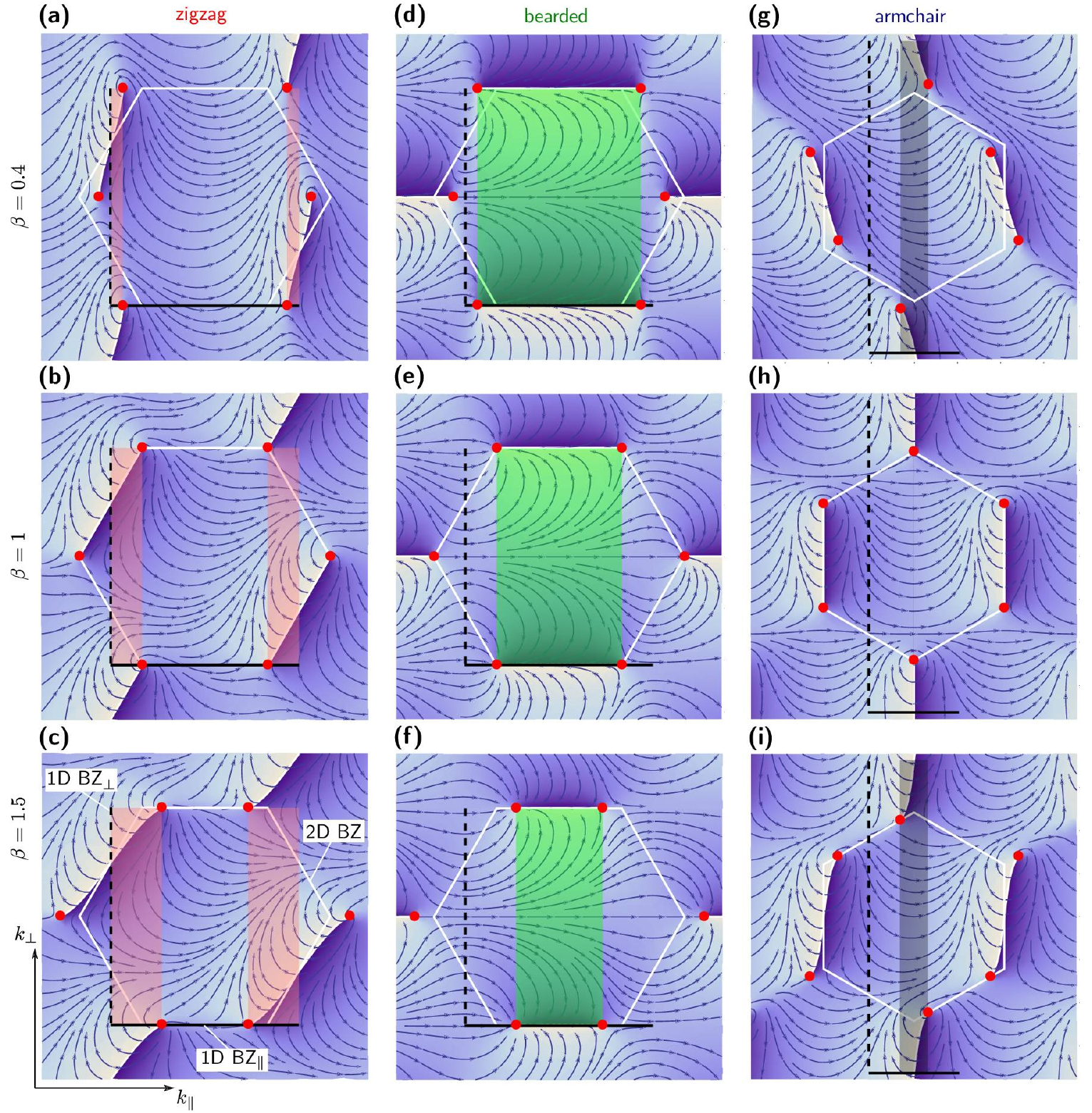}
	\caption{\label{Fig_Zak} Stream and density plots of the phase $\phi(k_\parallel,k_\perp)$ in reciprocal space, for the three edges (zigzag, bearded and armchai) and for various deformations (see the sketches in figures~\ref{Fig_DiagBZ} and~\ref{Fig_DiagAC}). The first 2D (bulk) BZ is indicated by a white contour. The red dots are the positions of the Dirac points. The first 1D BZ associated to the ($\parallel$) and ($\perp$) directions are shown by solid and dashed black lines, respectively. For a given $k_\parallel$, the existence of edge states is related to the winding of the phase along the $k_\perp$ direction in the BZ: there are edge states when the total rotation of the phase is $2\pi$ as seen in the shaded areas [${\cal Z}(k_\parallel) = \pi$]. There are no edge states otherwise [${\cal Z}(k_\parallel) = 0$].}
\end{figure}

Starting from the infinite unstrained system, in the $(A,B)$ basis, the effective Hamiltonian has the general form
\begin{equation} 
{\cal{H}}_{\k}= - t \left(
\begin{array}{cc}
0 & f^*(\k) \\
f(\k) & 0 \\
\end{array}
\right), \label{GH}
\end{equation}
where the function $f(\k)$ describes the coupling between atoms of one sublattice with the three nearest neighbors belonging to the other sublattice (see for instance \cite{Montambaux09}). The wavefunctions have the form
\begin{equation}
\psi_{\k}(\r) = {1 \over \sqrt{2} } \left(
\begin{array}{c}
\pm 1\\
e^{ i \phi(\k)}\\
\end{array}
\right)e^{i \k \cdot \r},
\end{equation}
where $\phi(\k)=\mbox{arg}[f(\k)]$ and $\pm$ corresponds to positive and negative energies. The winding of the relative phase $\phi(\k)$ in the reciprocal space has quite interesting properties. First of all, as seen on figure~\ref{Fig_Zak}, around each Dirac point, the phase rotates by $\pm 2 \pi$. The circulation of $\phi(\k)$ along a surrounding closed path is quantized: the Berry phase is defined by ${1/2}\oint d\k \, \nabla_{\k} \phi(\k) = \pm \pi$~\cite{CastroNeto09}. Our purpose here is to stress that the winding of the $\phi(\k)$ phase in reciprocal space carries an additional information related to the existence of edge states. Let us consider a ribbon geometry and define the directions parallel ($\parallel$) and perpendicular ($\perp)$ to the ribbon length. For a ribbon of finite width containing $M$ dimers, the perpendicular wavevector of a bulk state $k_\perp$ is quantized. For a two-component wavefunction, the quantization condition reads $k_\perp (M+1) a - \phi(\k)= \kappa \pi$ with $\kappa = 1, \cdots, M$~\cite{Delplace11}\footnote{Note that the quantization for a one-component wavefunction would be $k_\perp (M+1) a = \kappa \pi$.}.
This equation has $M$ or ($M-1$) roots depending on the winding of the phase under variation of $k_\perp$. When only ($M-1$) solutions exist, the missing solution corresponds to an edge state. More precisely one has the correspondence
\begin{equation*}
{\cal Z}(k_\parallel) = \left\{
\begin{array}{cl}
\pi & \longleftrightarrow \mbox{edge states}\\
 0  & \longleftrightarrow \mbox{no edge states} 
\end{array}
\right. \ ,
\end{equation*}
where the Zak phase $ {\cal Z}(k_\parallel)$ is the phase accumulated in the first 1D Brillouin zone along the $k_\perp$ direction:
\begin{equation}
{\cal Z}(k_\parallel) = {1 \over 2} \int_{BZ} d k_\perp {\partial \phi(k_\parallel,k_\perp) \over \partial k_\perp} \ .
\end{equation}

\begin{figure}[t]
	\centering
	\includegraphics[width=6cm]{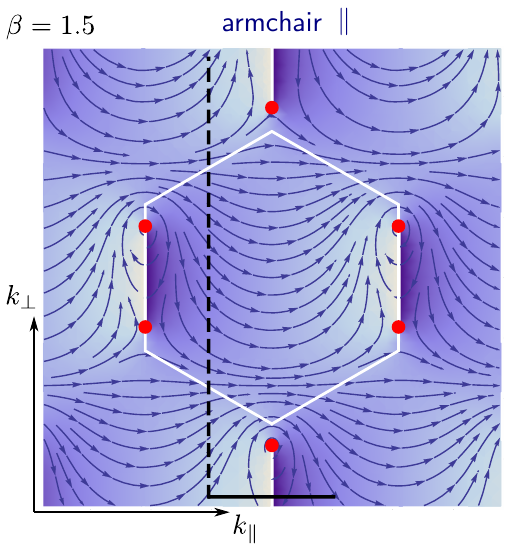}
	\caption{\label{Fig_ZakACP} Stream and density plots of the phase $\phi(k_\parallel,k_\perp)$ in reciprocal space, for the armchair edge and a deformation parallel to the edge. Marker and line legends are the same as in figure~\ref{Fig_Zak}. The Zak phase is $0$ everywhere, confirming the absence of edge state in this case.}
\end{figure}

Therefore existence of edge states may be read immediately from a plot of the phase $\phi(k_\parallel,k_\perp)$ as plotted in figure~\ref{Fig_Zak}.
For a given $k_\parallel$, if the rotation of the phase is $2 \pi$, there is an edge state, if the total rotation is $0$, there is no edge state. 
The function $\phi(\k)$ depends on the boundary. 
For a given type of edge, the ribbon is constructed by the translations of an elementary dimer (dotted box in figures~\ref{Fig_DiagBZ}(a),~\ref{Fig_DiagBZ}(b), and~\ref{Fig_DiagAC}(a)), so that the writing of the bulk Hamiltonian has a form which depends on the considered edge~\cite{Ryu02, Delplace11}.

Let us consider first the zigzag edge. The function $f(\k)$ reads in this case
\begin{equation}
f^{\mathrm{zz}}(\k)= 1 + \beta e^{i \k\cdot\a_1} + e^{i \k\cdot\a_2} = 1 + \beta e^{i ({\sqrt{3} \over 2} k_\parallel - {3\over 2} k_\perp)a}+ e^{i\sqrt{3}k_\parallel a} \ ,
\end{equation}
where the two elementary vectors $\a_1$ and $\a_2$ are shown on figure~\ref{Fig_DiagBZ}(a).
Similarly, the functions $f(\k)$ for the bearded (figure~\ref{Fig_DiagBZ}(b)) and armchair (figure~\ref{Fig_DiagAC}(a)) cases read, respectively
\begin{equation}
f^{\mathrm{bd}}(\k)= \beta + e^{i \k\cdot\a_1} + e^{i \k\cdot\a_2} = \beta + e^{i({\sqrt{3}\over 2} k_\parallel+{3 \over 2} k_\perp)a} + e^{i(-{\sqrt{3}\over 2} k_\parallel+{3 \over 2} k_\perp)a} \ ,
\end{equation}
\begin{equation}
f^{\mathrm{ac}}(\k)= 1 + \beta e^{i \k\cdot\a_1} + e^{i \k\cdot\a_2} = 1 + \beta e^{i({3 \over 2} k_\parallel-{\sqrt{3} \over 2} k_\perp)a} + e^{i({3 \over 2} k_\parallel+{\sqrt{3} \over 2} k_\perp)a} \ . \label{f-armchair}
\end{equation}
Figure~\ref{Fig_Zak} shows $\phi(\k)$ for the different edges and for various deformations characterized by the parameter $\beta$.
With increasing distortion, the density of edge states increases in the zigzag case, while it decreases in the bearded case as indicated by the red and green zone, respectively. In the armchair case, while there are no edge states in the undistorted lattice~\cite{Nakada96, Brey06, Kohmoto07}, they appear in the presence of a distortion which is not parallel to the edge. If the distortion is along the edge, the function $f(\k)$ reads $f^{\mathrm{ac}}(\k)= \beta + e^{i \k\cdot\a_1} + e^{i \k\cdot\a_2}$ instead of (\ref{f-armchair}). The corresponding plot of $\phi(k_{\parallel},k_{\perp})$ is shown on figure~\ref{Fig_ZakACP} and we see that ${\cal Z}(k_\parallel)=0$ everywhere, as all lines of phase jumps are parallel to $\k_{\perp}$, confirming the absence of edge states in this case. A thorough experimental investigation of armchair zero-modes is presented in the following section.

\section{Study of armchair edges in honeycomb lattice under uniaxial strain}
\label{Sec_ACstudy}

\begin{figure}[ht]
	\centering
	\includegraphics[scale=1]{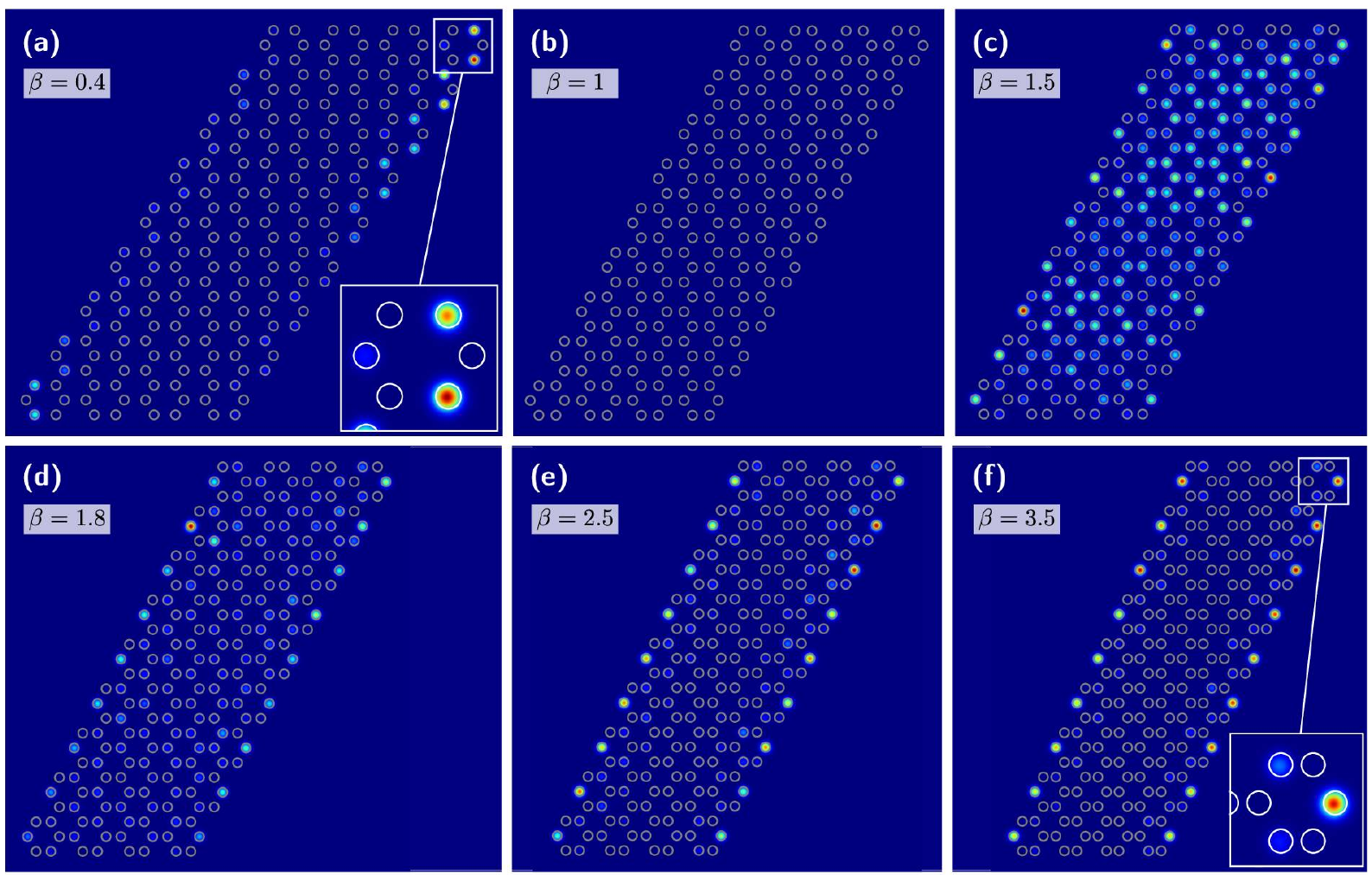}
	\caption{\label{Fig_WavefctAC} Experimental armchair zero-mode intensities for $\beta$ ranging from 0.4 to 3.5. (a) $\beta < 1$. (b) $\beta = 1$, no edge states. (c -- f) $\beta > 1$. Insets in (a) and (f) emphasize the switch between the two triangular sublattice excitation.}
\end{figure}

In previous work, we addressed the issue of topological phase transition in strained artificial graphene and incidentally observed armchair edge states~\cite{Bellec13a}. Here, we perform a quantitative study on ribbons with only armchair edges where the anisotropy axis is along their widths. Figure~\ref{Fig_WavefctAC} shows the zero-modes for $\beta$ ranging from 0.4 to 3.5. It is worth noting that no zero-modes appear on armchair edges along the anisotropy axis whatever the value of $\beta$. Moreover, for the case $\beta = 1$, the ribbon does not support any armchair edge states. These features were already observed with an hexagonal flake in \cite{Bellec13a} and are in accordance with the Zak phase analysis developed in section~\ref{Sec_Zak}. For $\beta \neq 1$, zero-modes along the two oblique edges are clearly present. Two main features have to be stressed: (\textit{i}) edge states live only on one of the two triangular sub-lattices. The excited sub-lattice depends on whether $\beta$ is less than or greater than 1, as seen in figures~\ref{Fig_WavefctAC}(a) and~\ref{Fig_WavefctAC}(c-f), respectively. The switch between the two sub-lattices is clearly visible in the insets. (\textit{ii}) The zero-mode localization length along the edges depends on the anisotropy and, for $\beta > 1$, decreases with $\beta$.

\begin{figure}[t]
	\centering
	\includegraphics[scale=1]{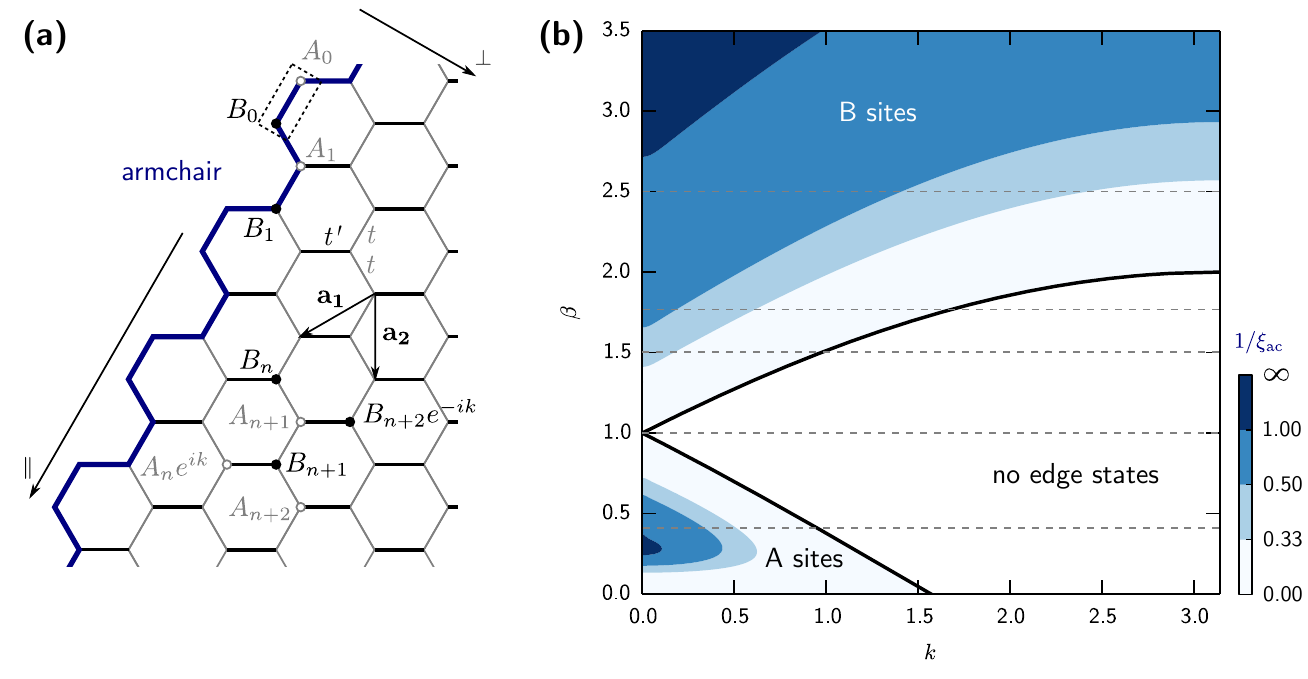}
	\caption{\label{Fig_DiagAC}(a) Schematic representation of a semi-infinite lattice with armchair edge along the $\parallel$ axis. The anisotropy axis is horizontal, the corresponding coupling strength is denoted $t'$. $A_n$ and $B_n$ are the wavefunction amplitudes in the two sublattices. $k$ is the Bloch wavevector (see text for details). (b) Armchair edge states existence diagram for various $k$ and $\beta$. The colormap corresponds to the inverse localization length $1/\xi_{ac}$. Light (dark) colors indicate a delocalized (localized) state. The white area indicates no edge states. For $\beta > 1$, only $B$-sites are excited. For $\beta < 1$, only $A$-sites are excited.}
\end{figure}

To understand the observations (\textit{i}) and (\textit{ii}), we propose a tight-binding analysis comparable to what has been reported in section~\ref{Sec_BZstudy}. Let us consider a semi-infinite honeycomb lattice with armchair edge (figure~\ref{Fig_DiagAC}(a)). The appropriate $A\!-\!B$ unit cell is represented by the dashed box. Now, the armchair edge is built by translating the dimer with vectors $m\,(\mathbf{a_1}+\mathbf{a_2})$ ($m$ integer). Here, $k \equiv 3k_{\parallel} a$, where $a$ is the lattice spacing and $k_{\parallel}$ is the 1D wave vector pertaining to the armchair edge. The 1D Brillouin zone is defined by $k_{\parallel} \in \left[ \frac{-\pi}{3a}, \frac{\pi}{3a} \right]$ corresponding to $k \in \left[ -\pi, \pi \right]$. The bulk lattice is built by translating the dimer with vectors $n\,\mathbf{a_2}$ ($n$ positive integer). The tight-binding recurrence equations for zero-modes read (see figure~\ref{Fig_DiagAC}(a)):
\begin{eqnarray}
A_{n+2} + A_{n+1} + \beta A_{n} e^{i k} & = & 0 \ , \label{Eq_A_AC} \\
B_{n+1} + B_{n} + \beta B_{n+2} e^{- i k} & = & 0 \ . \label{Eq_B_AC}
\end{eqnarray}
According to equation~(\ref{Eq_A_AC}), for $\beta>1$, the amplitude increases with $n$ so that $A$-sites cannot support any edge states, we have $A_n \equiv 0$, $\forall n$. Inversely, for $\beta<1$, equation~(\ref{Eq_B_AC}) implies $B_n \equiv 0$. Here, we have provided a simple explanation of the switching mechanism experimentally observed and described above.

For $\beta > 1$, with $B_0 = 1$ as initial condition and $B_1 = -e^{ik}/\beta$ to ensure $B_{i<0} \equiv 0$, the intensities on $B$-sites are given by
\begin{equation}
\label{Eq_IntBn_AC}
|B_n|^2 = |\lambda r_-^n + \mu r_+^n|^2 \equiv e^{-n/\xi_{\mathrm{ac}}} \ ,
\end{equation}
with
\begin{eqnarray}
r_{\pm} &=& \frac{-1 \pm \sqrt{1- 4 \beta e^{-ik}}}{2 \beta e^{-ik}} \ , \label{Eq_r_AC} \\
\lambda &=& \mu^* = \frac{B_0 r_+ - B_1}{r_+ - r_-} \ . \label{Eq_lambdaAC}
\end{eqnarray}

For $|r_+| <|r_-| < 1$, we obtain decaying solutions with
\begin{equation}
|k|< \arccos\left[\frac{\beta}{2}(3-\beta^2)\right] \ .
\label{Eq_existAC}
\end{equation}

\begin{figure}[t]
	\centering
	\includegraphics[scale=1]{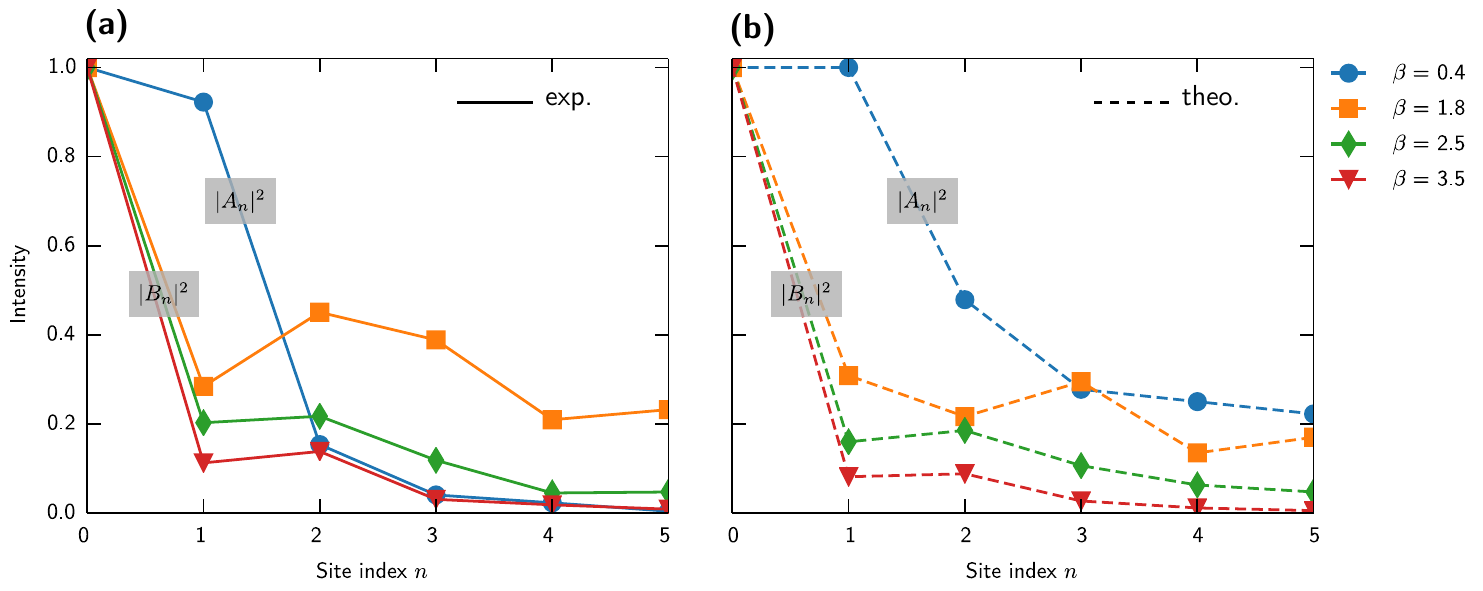}
	\caption{\label{Fig_ProfileAC} Experimental (a) and theoretical (b) armchair zero-mode intensity profiles for $\beta$ ranging from 0.4 to 3.5. For $\beta < 1$ ($\beta>1$), $A$ ($B$) sites are considered.}
\end{figure}

The diagram of existence of armchair edge states, defined by condition~(\ref{Eq_existAC}), is plotted in figure~\ref{Fig_DiagAC}(b). The blue color scale gives the transverse spatial extension of the armchair zero-modes extracted from equation~(\ref{Eq_IntBn_AC}). First, as pointed out in section~\ref{Sec_Zak} and observed in figure~\ref{Fig_WavefctAC}(b) armchair edges along the parallel axis do not support any edge states for $\beta = 1$. Then, we see that their existence is not related to the topological transition observed in~\cite{Bellec13a}: zero-modes appear as soon as $\beta \neq 1$. In the case $\beta =0.4$, $A$-sites support zero-modes occupying a limited range of $k$, as given by equation~(\ref{Eq_existAC}), almost half of them being well localized (dark blue colors in the diagram). For $1 < \beta < \beta_c=2$, the edge states supported by the $B$-sites are still belonging to a finite range of $k$ and are mainly delocalized (light blue colors in the diagram), as observed in the figures~\ref{Fig_WavefctAC}(b) and (c). For $\beta > 2$, zero-modes run over the full 1D Brillouin zone and their localization lengths decrease with $\beta$. To be more quantitative, we plotted in figure~\ref{Fig_ProfileAC} the intensities of $B_n$ ($A_n$), for $\beta >1$ ($\beta <1$) and $n$ ranging from 0 to 5. Figure~\ref{Fig_ProfileAC}(a) shows the experimental profiles extracted from figure~\ref{Fig_WavefctAC} by integrating over vertical lines of sites. Figure~\ref{Fig_ProfileAC}(b) depicts the intensities calculated from equations~(\ref{Eq_IntBn_AC})--(\ref{Eq_lambdaAC}) with an integration of $k$ over all possible values defined by equation~(\ref{Eq_existAC}). For $\beta = 2.5$ (green diamonds) and $\beta = 3.5$ (red triangles), the agreement is very good. For $\beta = 0.4$ (blue circles) and $\beta = 1.8$ (orange squares), the difference between measured and calculated profiles is more important. To explain these discrepancies, it is worth mentioning that for $\beta > \beta_c$ zero-modes actually appear in the bandgap and are therefore clearly isolated in the DOS. When $\beta < \beta_c$, bulk states are very close to the Dirac point and spoil the zero-mode intensity extraction. Moreover, the calculations are performed with semi-infinite lattices whereas experiments use finite ribbons. Consequently, in the experiments, we have to take into account a discrete sampling of the 1D Brillouin zone, the precise $k$-values depending on the ribbon length. Thus, the discrete sum of such zero-modes having different localization lengths can lead to a larger overall extension compared to the continuous case. Here again, $\beta$ stands as control parameter allowing notably to switch edge states from one sublattice to the other.

\begin{figure}[t]
	\centering
	\includegraphics[scale=1.2]{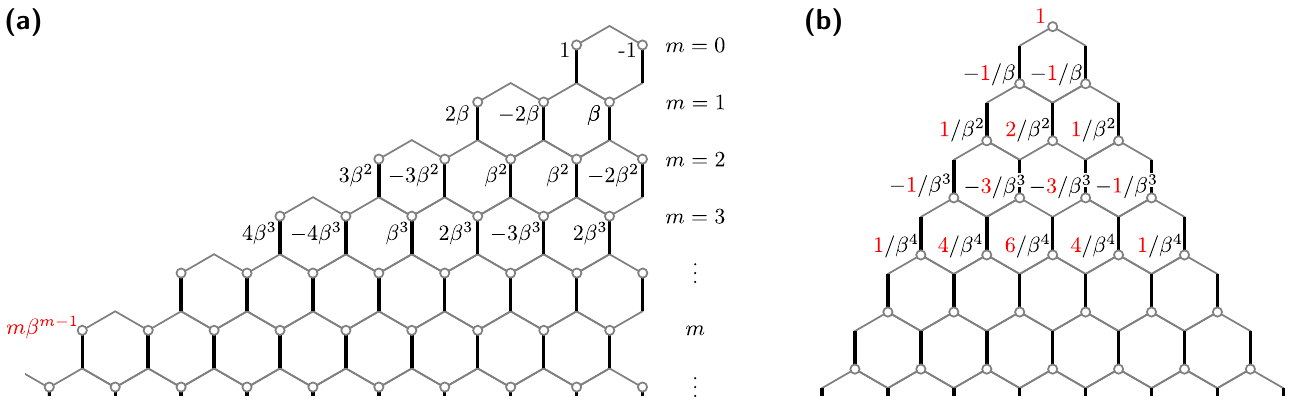}
	\caption{\label{Fig_CornerState} (a) Armchair zero-mode amplitude components for $\beta<1$ and $k=0$. The wavefunction is non-zero only on $A$-sites (empty circles). $m$ indicates the $A$-row index. The anisotropy axis is vertical, thick lines represent the corresponding coupling. (b) Zero-mode amplitude components for $\beta>2$ and $k=0$ appearing at the intersection of two zigzag edges.}
\end{figure}

To conclude this section, we would like to raise the existence of a new type of zero-modes in finite size systems located at the intersection of two edges, namely the \textit{corner state}. The comparison between figures~\ref{Fig_WavefctAC}(a) and~(f) clearly underlines that the distribution of the zero-mode along the parallel edge is homogeneous for $\beta = 3.5$ (i.e.\ all edge $B$-sites are illuminated) whereas the intensity decreases from the top-right corner for $\beta = 0.4$. Once more, a tight-binding analysis for zero-modes allows us to obtain non-homogeneous edge states. In figure~\ref{Fig_CornerState}(a), we build for $\beta < 1$ the amplitude of a $k=0$ edge state and obtain a decaying solution. Along the oblique edge, the amplitude reads $A_m = m \beta^{m-1}$, $m$ being the $A$-row index. Since localized edge states exist only for small $k$-values (dark blue zone in figure~\ref{Fig_DiagAC}(b)), this $k=0$ decaying solution dominates the overall intensity distribution. On the contrary, for $\beta = 3.5$, such decaying solutions may also exist for $k = 0$ but are compensated by uniform solutions associated with larger $k$-values, thus leading to an homogeneous overall intensity distribution.

One could expect that corner states exist for various type of edge intersections. For example, figure~\ref{Fig_CornerState}(b) indicates a corner state generated at the intersection of zigzag edges for $\beta >2$. The hierarchy of its amplitudes follows the rules of a Pascal triangle. The same hierarchy can be observed in figure~\ref{Fig_WavefctBZ}(d) along the direction going diagonally from the corner along a zigzag axis to the bulk. Corner states present strong similarities with zero-modes associated to localized impurities in anisotropic graphene~\cite{Dutreix13} and will be the subject of further studies.

\section{Conclusion}

We have studied edge states in artificial microwave graphene ribbons under uniaxial strain. By directly imaging the zero-modes, we have shown how the anisotropy parameter allows us to create and manipulate edge states. Based on a nearest-neighbor tight-binding analysis and supported by topological arguments (Zak phase), we have drawn diagrams of existence and localization length variation for zigzag, bearded and armchair edge states in remarkable agreement with experiments. Our results show also that higher order nearest-neighbor couplings, inherently present in the experiments~\cite{Bellec13b}, have no significant impact on the edge states. We also discuss the existence of a new type of zero-mode at the intersection of two type of edges, namely the corner state, whose study is left for a future work.

\ack
The authors are thankful to Pierre Delplace for enlightening discussions concerning the Zak phase during the international workshop on `Artificial Graphene' held at the Centro Internacional de Ciencias A.C (Cuernavaca, Mexico) in 2013. This work was supported by the DFG via the Forschergruppe 760.

\end{document}